\documentclass[twocolumn,prl]{revtex4}%
\usepackage{amsfonts}
\usepackage{amsmath}
\usepackage{amssymb}
\usepackage{graphicx}
\usepackage{type1cm}
\usepackage{color}
\usepackage{comment}%
\providecommand{\U}[1]{\protect\rule{.1in}{.1in}}
\begin{document}
\title{Chirality-induced Phonon Dispersion in a Noncentrosymmetric Micropolar Crystal}
\author{J. Kishine$^{1,2}$, A. S. Ovchinnikov$^{3,4}$, and A. A. Tereshchenko$^{3}$}
\affiliation{$^{1}$ Division of Natural and Environmental Sciences, The Open University of
Japan, Chiba, 261-8586, Japan}
\affiliation{$^{2}$ Institute for Molecular Science, Okazaki, Aichi 444-8585, Japan}
\affiliation{$^{3}$ Institute of Natural Sciences and Mathematics, Ural Federal University,
Ekaterinburg, 620083, Russia}
\affiliation{$^{4}$ Institute of Metal Physics, Ural Division, Russian Academy of Sciences,
Ekaterinburg, 620219, Russia}
\date{\today}

\begin{abstract}
Features of the phonon spectrum of a chiral crystal are examined within the
micropolar elasticity theory. This formalism accounts for not only
translational micromotions of a medium but also rotational ones. It is found
that there appears the phonon band splitting depending on the
left/right-circular polarization in a purely phonon sector without invoking
any outside subsystem. The phonon spectrum reveals parity breaking while
preserving time-reversal symmetry, i.e. it possesses true chirality. We find
that hybridization of the micro-rotational and translational modes gives rise
to the acoustic phonon branch with a ``roton'' minimum reminiscent of the
elementary excitations in the superfluid helium-4. We argue that a mechanism
of this phenomena is in line with Nozi\`{e}res' reinterpretation [J. Low Temp.
Phys. \textbf{137}, 45 (2004)] of the rotons as a manifistation of an
incipient crystallization instability. We discuss a close analogy between the
translational and rotational micromotions in the micropolar elastic medium and
the Bogoliubov quasiparticles and gapful density fluctuations in ${}^{4}$He.

\end{abstract}

\maketitle

\textit{Introduction.}$-$ In hierarchy of electric, magnetic and mechanical
degrees of freedom and attendant interrelationships namely mechanical and
elastic properties of solids are most sensitive to structure. This is why the
mechanical response in solids has recently been the target of a new branch of
electronics, referred to as straintronics\cite{Bukharaev2018}. Connection of
structural chirality with static and dynamical properties offers a key to
understand functionality of chiral systems \cite{Barron2012}. A quintessential
example of the chirality-controlled phenomena is the optical
activity\cite{BarronBook2nd}, where propagation of circularly polarized light
through a chiral material depends on its handedness. As for elastic degrees of
freedom, a mechanical counterpart of optical activity, so-called acoustical
activity, has been attracting revived attention\cite{Frenzel2019}. This
phenomena was first predicted by Portigal and Burstein\cite{Portigal1968} and
direct observation thereof was subsequently provided for the $\alpha-$quartz
crystal belonging to an enantiomorphic space group\cite{Pine1971}.

\begin{comment}
The optical activty, as a
paradigm for chirality, is an example of true chirality according to Laurence
Barrons' definition\cite{BarronBook2nd}, i.e., true chirality is exhibited by
systems, where parity $\mathcal{P}$ is broken but not time reversal
$\mathcal{T}$ combined with any proper rotation $\mathcal{R}$. This modified
definition is quite natural in the sense that the dynamical meaning of
chirality is a coupling of rotation and translation, i.e. it is an equivalent
of helicity. The optical activity is also known to result from a nonlocal
response of a crystal to a light wave, when there are first-order spatial
dispersion contributions to the dielectric constant determined via the optical
gyrotropic tensor.
%\cite{Rukhadze1961}%
\end{comment}

The acoustical activity is also related with first-order spatial dispersion
contributions to the elastic constants \cite{Portigal1968}. This situation is
somewhat similar to a role of Dzyaloshinskii-Moriya interaction which leads to
a linear Lifshitz invariant in free energy of a chiral helimagnet
\cite{Book2015,Paterson2020}. However, chirality effects are beyond the
conventional elasticity theory \cite{Landau}, which considers only a local
translation of points and the force stress (force per unit area) but
completely ignores a local rotation of these points and the concominant couple
stress (a torque per unit area).
{As a consequence, there appears the elastic
four-rank tensor, $C_{ijkl}$, which connects the parity-even and odd 2nd rank
tensors, and gives rise to the chiral term in the energy functional
($C_{klmn}\varepsilon_{kl}\gamma_{mn}$ as explained below).}

\begin{comment} This theory accomodates size effects in
elastic behaviour, which are expected to appear as the consequence of the
largest structural elements in solids, and allowed for consideration of
chirality effects in the elastic media. The latter have recently become topics
for study in metamaterials\cite{Frenzel2017} and in quasi two-dimensional
monolayers of noncentrosymmetric tungsten diselenide crystals\cite{Zhu2018}.
It must be recognized, however, that applications of the micropolar elastisity
theory to noncentrosymmetric materials have been mainly limited to mechanical
engeneering problems, including investigations of static deformations in bones
and synthetic composites containing twisted fibers\cite{Lakes1982}, an
analysis of static and wave properties of tetrachiral lightweight
lattices\cite{Chen2014}.
\end{comment}

These missing effects, which may be viewed as a perticular manifestation of
nonlocality, are addressed in the micropolar elasticity
theory\cite{Eringen1999,Nowacki1986}. So far, only few attempts have been made
to calculate dispersion curves of the micropolar elastic waves in crystals. We
note in this regard the pioneering research undertaken by Pouget at al. for
the centrosymmetric compound KNO$_{3}$\cite{Pouget1986}, where spectrum of the
micropolar waves reciprocal in the momentum space was obtained.
\begin{comment}, they presented a lattice
theory of ferroelectric crystals with molecular groups, which accounts for
their microrotations added to the usual displacements. In the long-wavelength
limit this approach coincides with the micropolar theory of elasticity.
Dispersion curves demonstrate mixing of the longitudinal and transverse
acoustic modes with the rotational one, but the spectrum remains symmetric
against the sign of crystal momentum.
\end{comment}

Then, the natural question arises as to whether a phonon spectrum in a a
chiral crystal exhibits nonreciprocity effects in propagation of micropolar
elastic waves. This is the question addressed in this Letter. A salient
feature of the nonreciprocity is a polarization-dependent splitting of phonon
bands similar to the one for electronic bands due to spin-orbit
coupling\cite{Rashba1959}. In contrast to the phonon magneto-chiral effect
\cite{Tereshchenko2018,Nomura2019}, for which the splitting is achieved
through coupling with nonreciprocal magnons, our aim is to find the result for
a purely phononic sector without the involvement of any subsystem outside.

Another issue discussed in this Letter is how the structural chirality relates
to the phonon angular momentum and spin. In this regard, we mention the recent
studies of chiral phonons in monolayers of hexagonal \cite{Zhang2015} and
kagome lattices \cite{Chen2019}. In these systems, the phonon eigenmodes at
high-symmetry points of the Brillouin zone (BZ) inherit the threefold
rotational symmetry of the lattice which allows to label these phonon
eigenmodes with pseudoangular momentum. It includes both orbital and spin
parts, the latter coincides with the phonon chirality characterized by the
circular polarization of phonons. In contrast to this scenario in which
chirality is assigned only to special BZ points, we examine an effect of the
structural chirality on phonon dispersion over the whole Brillouin zone. The
chiral helimagnet CrNb$_{3} $S$_{6}$ serves to illustrate our results, one of
which is striking similarity of microrotations embedded in the micropolar
elastisity theory with roton excitations in helium-4.

\begin{comment} \textit{General theory.}$-$ We here briefly outline a general scheme of
treatment of elementary excitations in the context of the micropolar
elasticity theory. All technical details may be found in Supplemental
Material\cite{Supplement}.
\end{comment}

\begin{figure}[t]
\begin{center}
\includegraphics[width=80mm]{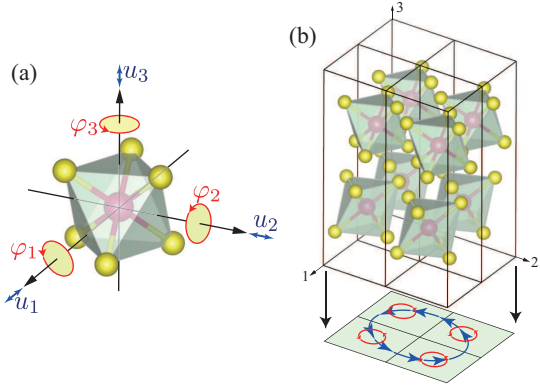}
\end{center}
\caption{(a) Schematic view of the rotational and translational degrees of
freedom of the atomic microelement. As an example of the microelement, we show
the CrS$_{6}$ block inside the elementary cell of CrNb$_{3}$S$_{6}$. (b)
Typical distribution of microrotation fields. We also depict conceptual
representation of the micro-rotation (red small circle with arrows) and
translation (blue large circle with arrows) associated with a circularly
polarized mode, which are projected on the $1-2$ plane. }%
\label{model}%
\end{figure}

\textit{Chiral phonon dispersions for the point group $622$.}$-$ A general
scheme of treatment of elementary excitations in the context of the micropolar
elasticity theory may be found in Supplemental Material\cite{Supplement}. In
this theory, the field of translational displacements, $\boldsymbol{u}\left(
\boldsymbol{r}\right)  $, is supplemented by the field of microrotations,
$\boldsymbol{\varphi}\left(  \boldsymbol{r}\right)  $, and the both are
attributed to a microelement located at the position $\boldsymbol{r}$ (see
Fig. \ref{model}). Two linear {micropolar} strain tensors
$\varepsilon_{kl}=\partial_{l}u_{k}-\epsilon_{klm}\varphi_{m}$ and
$\gamma_{kl}=\partial_{l}\varphi_{k}$ form measures of microdeformations. Here
and throughout, $\epsilon_{klm}$ is the Levi-Civita symbol and Einstein
summation convention is used. The strain energy density is given by the
quadratic form, $U=\frac{1}{2}A_{klmn}\varepsilon_{kl}\varepsilon_{mn}%
+\frac{1}{2}B_{klmn}\gamma_{kl}\gamma_{mn}+C_{klmn}\varepsilon_{kl}\gamma
_{mn}$, where
{the third term (chiral term) changes its sign under the
inversion operation while the first and the second do not. This chiral coupling leads to the left/right-circular-polarization-depdendent non-reciprocal phonon dispersion.}

To demonstrate how the chiral coupling gives rise to new peculiarities of
phonon dispersion, we consider, as an example, the layered compound CrNb$_{3}%
$S$_{6}$ which has the non-centrosymmetric hexagonal space group $P6_{3}22$.
In this material, the Cr atoms are intercalated between the sandwich layers
S-Nb-S of the disulfide NbS$_{2}$ and surrounded by the six S atoms in an
octahedral geometry. The CrS$_{6}$ octahedra are not linked to each other. The
distance $d$(Cr-S)$=$ $2.393\mathring{A}$\cite{Volkova2014} is less than
$\sim0.1\mathring{A}$ the distance d(Nb-S)= $2.47\div2.50$%
\AA \ \cite{Sarode1986}. Assuming that modes due to relative displacements of
the Cr and S ions are not excited, the CrS$_{6}$ may be modeled as a rigid
structural unit. Below, we consider plane-waves propagating along the the
crystalline $[001]$ axis, i.e. the chiral axis.

Dispersion of the micropolar waves is obtained in the following way. To
describe transverse modes, it is appropriate to introduce the circular basis,
$u_{\pm}=u_{1}\pm iu_{2},$ and $\varphi_{\pm}=\varphi_{1}\pm i\varphi_{2}$,
where $+$/$-$ corresponds to the left/right circularly polarized
microdeformation fields. \begin{comment}
For the $622$ point group, the polar tensors
$\mathbf{A}$, $\mathbf{B}$ have 8 independent elements, while the axial one
$\mathbf{C}$ does 10 (see, for instance, Tables D.20 and D.22 in Ref.
\cite{Sirotin1982}).
\end{comment}
Then, the EOMs for the decoupled transverse modes may be written as,
\begin{align}
\rho\ddot{u}_{\pm}  &  =A_{55}\partial_{3}^{2}u_{\pm}+C_{74}\partial_{3}%
^{2}\varphi_{\pm}\mp i\left(  A_{47}-A_{55}\right)  \partial_{3}\varphi_{\pm
},\label{EOMutransverse}\\
\rho j_{\pm}\ddot{\varphi}_{\pm}  &  =C_{74}\partial_{3}^{2}u_{\pm}\mp
i\left(  A_{47}-A_{55}\right)  \partial_{3}u_{\pm}+B_{44}\partial_{3}%
^{2}\varphi_{\pm}\nonumber\\
&  \mp2i\left(  C_{44}-C_{74}\right)  \partial_{3}\varphi_{\pm}-\left(
A_{44}-2A_{47}+A_{55}\right)  \varphi_{\pm}, \label{EOMphitransverse}%
\end{align}
where $\rho=5.029$ g/cm$^{3}$\cite{Ghimire2013} and $j_{\pm}=j_{11}%
=0.5\times10^{-19}$ m${}^{2}$\cite{MicroInertia} are the mass density and the
microinertia tensor component, respectively. Here, the four-rank tensor
elements are represented in the Sirotin (generalized
Voigt)\ scheme\cite{Supplement}. The most important point is the appearance of
the linear gradient terms, $\partial_{3}u_{\pm}$ and $\partial_{3}\varphi
_{\pm}$, which cause polarization-dependent velocities. Hybridization of these
circularly polarized $u_{\pm}$- and $\varphi_{\pm}$-modes gives rise to
acoustic and optical branches of the spectrum of propagating transverse waves
in the micropolar medium. It is to be noted that each of the modes, $u_{\pm}$
(or $\varphi_{\pm}$), is non-reciprocal in real space, but they form a pair
invariant under time-reversal symmetry. This may be inferred from Eqs.
(\ref{EOMutransverse}) and (\ref{EOMphitransverse}) which break parity
$\mathcal{P}$ but preserve time-reversal symmetry $\mathcal{T}$. This
situation means that phonons in the micropolar chiral crystal exhibit
\textit{true chirality} in contrast to the
phonon\cite{Tereshchenko2018,Nomura2019} or electrical\cite{Aoki2019}
magneto-chiral effects, where both $\mathcal{P}$ and $\mathcal{T}$ are
simultaneously broken. On that understanding, we call these excitations truly
chiral phonons. This polarization dependent splitting of the transverse phonon
branches is analogous to the Rashba splitting of electronic bands, in which
$\mathcal{P}$ is also broken but not $\mathcal{T}$. Nonreciprocity of the
truly chiral phonons originates from the $\left(  C_{44}-C_{74}\right)
$-coupling term in the EOMs (\ref{EOMutransverse},\ref{EOMphitransverse}) and
vanishes for $C_{44}=C_{74}$. \begin{comment}
This chiral
coupling between microtranslations and microdeformations appears in the energy
functional as $\left(  C_{44}\varepsilon_{\alpha3}+C_{74}\varepsilon_{3\alpha
}\right)  \gamma_{\alpha3}$, $\alpha=1,2$.
\end{comment}

\begin{figure}[ptb]
\begin{center}
\includegraphics[width=85mm]{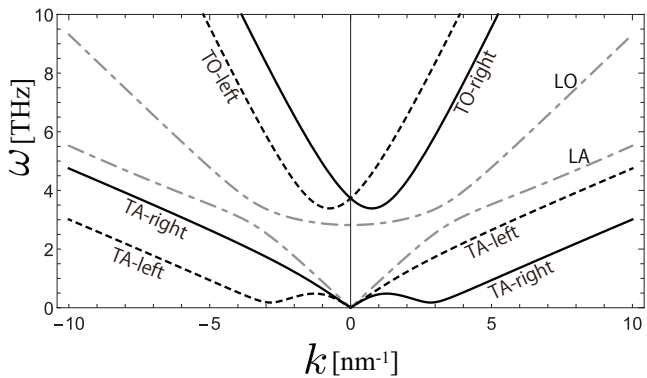}
\end{center}
\caption{ Phonon dispersion curves for the chiral micropolar crystal: the
longitudinal acoustic (LA) and optical (LO) branches (dashed-and-dotted line),
the transverse left-handed acoustic (TA-left) and optical (TO-left) branches
(dashed line), the transverse right-handed acoustic (TA-right) and optical
(TO-right)\ branches (solid line). Numerical values of the tensor components
are chosen as $A_{33}=0.4\cdot10^{10}$ N/${\text{m}}^{2}$, $A_{66}%
=4.9\cdot10^{10}$ N/${\text{m}}^{2}$, $A_{69}=4.7\cdot10^{10}$ N/${\text{m}%
}^{2}$, $B_{33}=1.5\cdot10^{-10}$ N, $C_{33}=0.3$ N/m for the longitudinal
modes and $A_{44}=0.21\cdot10^{10}$ N/${\text{m}}^{2}$, $A_{55}=0.215\cdot
10^{10}$ N/${\text{m}}^{2}$, $A_{47}=0.195\cdot10^{10}$ N/${\text{m}}^{2}$,
$B_{44}=1.0\cdot10^{-10}$ N, $C_{44}=0.44$ N/m, $C_{74}=0.36$ N/m for the
transverse modes, respectively.}%
\label{SpectrumAll}%
\end{figure}

The equations that govern propagation of the longitudinal branches are written
as
\begin{align}
\rho\ddot{u}_{3}  &  =A_{33}\partial_{3}^{2}u_{3}+C_{33}\partial_{3}%
^{2}\varphi_{3},\label{EOMulong}\\
\rho j_{3}\ddot{\varphi}_{3}  &  =C_{33}\partial_{3}^{2}u_{3}+B_{33}%
\partial_{3}^{2}\varphi_{3}-2\left(  A_{66}-A_{69}\right)  \varphi_{3},
\label{EOMphilong}%
\end{align}
where $j_{3}=j_{33}$ will be taken as $1.0\times10^{-19}$ m${}^{2}$. Note the
absense of linear gradient terms that entails simple hybridization of the
$u_{3}$ and $\varphi_{3}$ modes.

Phonon dispersion relations may be obtained inserting plane waves propagating
along the chiral ($x_{3}$) axis, $u_{\alpha}\left(  x_{3},t\right)
=u_{\alpha}e^{i\left(  kx_{3}-\omega t\right)  }$ and $\varphi_{\alpha}\left(
x_{3},t\right)  =\varphi_{\alpha}e^{i\left(  kx_{3}-\omega t\right)  }$, into
Eqs. (\ref{EOMutransverse}-\ref{EOMphilong}). Hereinafter, the index $\alpha$
labels either the transverse-left/right ($"+/-"$) circular or longitudinal
($"3"$) polarization of the phonon branches. The results are summarized in
Fig. \ref{SpectrumAll}. There are six branches in total, namely, we have the
longitudinal acoustic (LA)/optical (LO), the transverse left-handed acoustic
(TA-left)/optical (TO-left), and the transverse right-handed acoustic
(TA-right)/optical (TO-right) modes. The associated dispersion relations take
the form
\begin{equation}
\left[  \omega_{\alpha}^{(\text{O/A})}\right]  ^{2}=\frac{1}{2\rho j_{\alpha}%
}\bigg[b_{\alpha}+j_{\alpha}a_{\alpha}\pm\sqrt{\left(  b_{\alpha}-j_{\alpha
}a_{\alpha}\right)  ^{2}+4j_{\alpha}\Delta_{\alpha}^{2}}\bigg].
\label{MPTspectrum}%
\end{equation}
The upper and lower $\pm$ signs in the r.h.s. of Eq.(\ref{MPTspectrum})
correspond to the optical (O) and acoustic (A) branches, respectively. The
$k$-dependent parameters ($a_{\alpha}$, $b_{\alpha}$, and $\Delta_{\alpha}$)
are given by $a_{\pm}=A_{55}k^{2}$, $b_{\pm}=B_{44}k^{2}\mp2\left(
C_{44}-C_{74}\right)  k+\left(  A_{44}+A_{55}-2A_{47}\right)  $, $\Delta_{\pm
}=C_{74}k^{2}\mp\left(  A_{47}-A_{55}\right)  k,$ and $a_{3}=A_{33}k^{2}$,
$b_{3}=B_{33}k^{2}+2\left(  A_{66}-A_{69}\right)  $, $\Delta_{3}=C_{33}k^{2}$.
In the long wavelength limit, $k\rightarrow0$, the frequencies of the acoustic
branches are proportional to the wavenumber, while the frequencies of the
optical branches tend to finite values, $\omega_{\pm}^{\text{(O)}}\left(
0\right)  =\sqrt{\left(  A_{44}+A_{55}-2A_{47}\right)  /\rho j_{\pm}}$ and
$\omega_{3}^{\text{(O)}}\left(  0\right)  =\sqrt{\left(  A_{66}-A_{69}\right)
/\rho j_{3}}$.

\begin{figure}[ptb]
\begin{center}
\includegraphics[width=45mm]{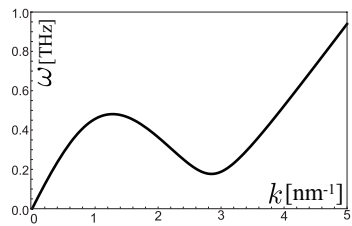}
\end{center}
\caption{The transverse right-handed acoustic (TA-right) branch that exhibits
the roton-like minimum. }%
\label{rotonlike}%
\end{figure}

\textit{Similarity to roton spectrum.}$-$ Figure \ref{rotonlike} contains only
the TA-right handed phonon mode specially selected from all branches shown in
Fig. \ref{SpectrumAll}. We observe that hybridization of the rotational and
translational modes gives rise to the lowest phonon (TA-right or TA-left)
branch which exhibits a \textquotedblleft roton minimum\textquotedblright%
\ around $k_{m}\sim\left(  C_{44}-C_{74}\right)  /B_{44},$\ reminiscent of the
excitation spectrum in superfluid${}^{4}$He \cite{Landau1941}. The minimum
occurs around a scale inversely proportional to the unit cell length and
reflects hybridization of the rotational and translational degrees of freedom
of the microelement CrS$_{6}$.

\begin{figure}[ptb]
\begin{center}
\includegraphics[width=60mm]{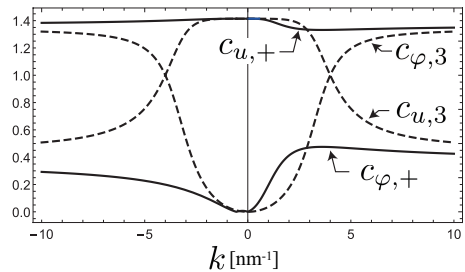}
\end{center}
\caption{The wavenumber dependence of the weights of translational and
rotational modes in the hybridized TA-right (solid line) and LA (solid line)
branches. \ }%
\label{Weights}%
\end{figure}

To elucidate reasons behind an appearance of the roton-like minimum, we trace
how translational and rotational modes contribute to the hybridized
excitations. We find weights of these modes specified by the dimensionless
coefficients $c_{u,\alpha}=u_{\alpha}/\sqrt{j_{\alpha}}$ and $c_{\varphi
,\alpha}=\varphi_{\alpha}$, whose explicit expressions are reproduced in
Supplemental Material\cite{Supplement}. Their $k$-dependence is illustrated by
Fig. \ref{Weights} for the TA-right and LA branches. It is evident that the
$u_{+}$-$\varphi_{+}$ hybridization causes resonant enhancement of the
rotational degrees of freedom ($\varphi_{+}$) when approaching the roton
minimum. In contrast, the ratio between the coefficients $c_{u,3}$ and
$c_{\varphi,3}$ is reversed near the crossing of the corresponding dispersion
curves, which are modified into the hybridized LA and LO phonon modes.

There is actually an interesting parallel between the TA chiral phonons and
roton excitations in superfluid ${}^{4}$He. Originally, the rotons had been
interpreted as a signiture of some local vorticity. According to Feynman's
view\cite{Landau1941}, a maximum in the static form factor $S_{\boldsymbol{q}%
},$ which signals a short range crystalline order, gives a roton minimum in
the excitation spectrum through the relation $\omega_{\boldsymbol{q}}%
=q^{2}/2mS_{\boldsymbol{q}}$. Later, Nozi\`{e}res\cite{Nozieres,Nozieres2}
proposed an alternate scenario, where the rotons should be viewed as an
incipient soft mode associated with a crystallization instability. According
to this view, the Bogoliubov quasiparticles hybridize with soft density
fluctuations to cause a resultant roton minimum. The picture can be formulated
by assuming two coupled excitations where the first correspond to the
Bogoliubov quasiparticles with the spectrum $\varepsilon_{\boldsymbol{q}}%
^{B}=\left[  \xi_{\boldsymbol{q}}^{2}+2\xi_{\boldsymbol{q}}N_{0}U\right]
^{\frac{1}{2}}$ and the second do to the density fluctuation mode, which is
characterized by the single mode frequency $\Omega_{\boldsymbol{q}}%
$\cite{Nozieres,Nozieres2}. Here, $\xi_{\boldsymbol{q}}$ is the boson kinetic
energy, $N_{0}$ is a condensate fraction, and $U$ is a direct repulsion
between bosons. The spectrum of the hybridized quasiparticles has the form
similar to Eq. (\ref{MPTspectrum}),
\begin{align}
E_{\boldsymbol{q}}^{2}  &  =\frac{1}{2}\left(  \Omega_{\boldsymbol{q}}%
^{2}+\left(  \varepsilon_{\boldsymbol{q}}^{B}\right)  ^{2}\right) \nonumber\\
&  \pm\frac{1}{2}\left[  \left(  \Omega_{\boldsymbol{q}}^{2}-\left(
\varepsilon_{\boldsymbol{q}}^{B}\right)  ^{2}\right)  ^{2}+16\alpha
_{\boldsymbol{q}}^{2}\xi_{\boldsymbol{q}}\Omega_{\boldsymbol{q}}\right]
^{\frac{1}{2}}, \label{BLspectrum}%
\end{align}
where $\alpha_{\boldsymbol{q}}$ is a strength of coupling between the bosons
and density fluctuations. Direct comparison between Eqs.(\ref{MPTspectrum})
and (\ref{BLspectrum}) eloquently illustrates the one-to-one correspondence
between the translational displacement, $\boldsymbol{u}\left(  \boldsymbol{r}%
\right)  $, and the microrotation, $\boldsymbol{\varphi}\left(  \boldsymbol{r}%
\right)  $, on one hand and the Bogoliubov quasiparticles and density
fluctuations on the other [$a_{\pm}(k)$ and $b_{\pm}(k)$ respectively
correspond to $\varepsilon_{\boldsymbol{q}}^{B}$ and $\Omega_{\boldsymbol{q}}$].

However, the roton minimum of the superfluid helium arises in the spectrum of
the \textit{longitudinal} sound wave of the normal component, while in the
micropolar crystal this effect is observed in propagation of the
\textit{transversal} elastic waves, but not for the longitudinal ones. This
difference may materialize arguments propounded by Landau and Feynman that the
rotons are related to local vorticity \cite{Landau1941}. The micro-rotation
does warrant the name roton as spinning motion of the microelement.

\begin{figure}[ptb]
\begin{center}
\includegraphics[width=65mm]{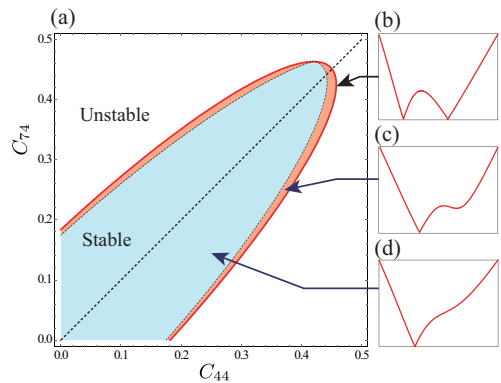}
\end{center}
\caption{(a) Phase diagram in the $C_{44}$-$C_{74}$ plane. The coloured area
corresponds to stability of the crystal structure. The roton minimum appears
in the vicinity of the soft mode instability (red). The concomitant profiles
of the spectrum are shown on the right (b-d). Numerical parameters are the
same as in Fig. \ref{SpectrumAll}.}%
\label{phase-diagram}%
\end{figure}

There is further similarity between the rotons in${}^{4}$He and the chiral
phonons. The boson condensate lowers the roton minimum but the latter will
remain finite due to the effect of condensate
depletion\cite{Nozieres,Nozieres2}. For the chiral phonons, a finite value of
the roton minimum may be inferred from the requirement that the crystal must
be stable against propagating elastic waves. In in Supplemental
Material\cite{Supplement}, we give proof that the stability condition implies
that the roton minimum will remain finite (never touch zero). An area of
stabilty within the $C_{44}$-$C_{74}$ plane (these constants responsible for
parity breaking) is plotted in Fig. \ref{phase-diagram}, together with a
region where the roton minimum emerges.

{The roton minimum occurs around $k\sim3$\,nm$^{-1}$ roughly corresponds to 20 nm in real space, so the continuum theory developed here may be still be valid. The inelastic neutron or Brillouin light scattering measurements may potentially probe  this finite $k$ excitation.
More quantitative arguments may require to build a
bridge between the present continuum theory and atomic lattice model, which is left for future studies.}
\begin{comment} On the other hand, taking account of the lattice constant $c= 12.12$ \AA, the minimum is located almost  halfway to the Brillouin zone edge. In the present letter, we used the continuum limit and the information on the Brillouin zone is missing. To extend the present scheme to the atomic lattice case is left for future research.
\end{comment}

\textit{Phonon angular momentum.}$-$ Once the spectrum of the micropolar waves
is known, one is capable to determine naturally orbital and spin parts of the
angular momentum associated with elastic deformations. By specifying the basis
of the left-handed ($L$) and right-handed ($R$) circularly polarized
transverse modes as $|R_{u}\rangle=|L_{u}\rangle^{\ast}=\frac{1}{\sqrt{2}%
}\left(  1,i,0,0,0,0\right)  ^{\text{T}}$, and $|R_{\varphi}\rangle
=|L_{\varphi}\rangle^{\ast}=\frac{1}{\sqrt{2}}\left(  0,0,0,1,i,0\right)
^{\text{T}}$ the solution $|\boldsymbol{v}\rangle^{T}=\left(  c_{u,1}%
,c_{u,2},c_{u,3},c_{\varphi,1},c_{\varphi,2},c_{\varphi,3}\right)  $ of Eqs.
(\ref{EOMutransverse}-\ref{EOMphilong}) may be decomposed as $|\boldsymbol{v}%
\rangle=\sum_{\alpha=u,\varphi}\langle R_{\alpha}|\boldsymbol{v}%
\rangle|R_{\alpha}\rangle+\langle L_{\alpha}|\boldsymbol{v}\rangle|L_{\alpha
}\rangle$. Then, the generator of rotations of the phonon polarization plane
around the $z$-axis, or phonon spin \cite{Zhang2015}, is $\hat{S}^{z}%
=\hbar\sum_{\alpha=u,\varphi}\left(  |R_{\alpha}\rangle\langle R_{\alpha
}|-|L_{\alpha}\rangle\langle L_{\alpha}|\right)  $ , so that
\begin{equation}
\langle\boldsymbol{v}|\hat{S}^{z}|\boldsymbol{v}\rangle=\frac{\hbar}{2}%
\sum_{f=u,\varphi}\left(  \left\vert c_{f,-}\right\vert ^{2}-\left\vert
c_{f,+}\right\vert ^{2}\right)
\end{equation}
is fulfilled. Obviously, the phonon circular polarization is quantized, i.e.
it can take only the $\pm\hbar$ values for the "$\mp$"-branches, respectively.
Unlike chiral phonons in the theories\cite{Zhang2015,Chen2019}, the phonon
spin is introduced for any $k$-point of the Brillouin zone. Neglecting the
microrotations $\boldsymbol{\varphi}$, the definition coincides with the spin
part of the phonon angular momentum given in Ref.\cite{Garanin2015},
$\boldsymbol{L}_{s}=\hbar\sum_{\boldsymbol{k}}{\boldsymbol{e}}_{\boldsymbol{k}%
}\left(  a_{\boldsymbol{k}-}^{\dagger}a_{\boldsymbol{k}-}-a_{\boldsymbol{k}%
+}^{\dagger}a_{\boldsymbol{k}+}\right)  $, where the angular momenta $\hbar$
of the individual $(\boldsymbol{k}\pm)$-phonons add up either parallel or
anti-parallel to their wave vectors, ${\boldsymbol{e}}_{\boldsymbol{k}%
}=\boldsymbol{k}/|\boldsymbol{k}|$. Classical interpretation of the resultant
$\boldsymbol{L}_{s}$ corresponds to small-radius circular shear displacements
of points around their equilibrium positions [Fig. \ref{model}(b)].

With regard to the orbital part of the phonon angular momentum, its
definition, $L_{\pm}=\rho j_{\pm}\dot{\varphi}_{\pm}$, follows directly from
the appropriate conservation law of the micropolar theory\cite{Eringen1999}.
This implies propagation of plane waves of the angular momentum density.

\textit{Roton and} \textit{acoustic activity.}$-$Nonreciprocity of the chiral
phonons results in consequential acoustic activity, most notably in the
vicinity of the roton minimum. By definition, the effect occurs when incident
transverse acoustic waves, which are linealy polarized, propagate in the
crystal along the $z$-axis. It is characterized by the rotation angle,
$\phi=\frac{1}{2}\omega l\left(  1/v_{-}-1/v_{+}\right)  $, at a distance $l$
from the incident surface. The phase velocities $v_{\pm}$ of the circularly
polarized transverse modes may be calculated from the dispersion relations
(\ref{MPTspectrum}) to give
\begin{align}
\frac{\phi}{l}  &  =\rho\omega^{2}\frac{\left(  A_{47}-A_{55}\right)
}{\left(  A_{44}A_{55}-A_{47}^{2}\right)  ^{2}}\nonumber\\
&  \times\left[  A_{55}C_{44}+A_{44}C_{74}-A_{47}\left(  C_{44}+C_{74}\right)
\right]  .
\end{align}
Comparison with EOMs (\ref{EOMutransverse},\ref{EOMphitransverse}) shows that
apart from the intercross coupling $A_{55}-A_{47}$ between the
microtranslations and microrotations, the inherent gyrotropy $C_{44}-C_{74}$
of the microrotations is an essential element of acoustic activity. Unlike the
micropolar theory, the conventional approach \cite{Portigal1968,Pine1971}
relates the latter to non-local interaction between stress and strain; this is
reflected in the first-order dispersion in expansion of the elastic
coefficients $c_{ij}(\boldsymbol{k},\omega)$. As a consequence, a difference
in phase velocities of circularly polarized waves appears at any
$\boldsymbol{k}$ vector.

\textit{Concluding remarks.}$-$We demonstrate polarization-dependent splitting
of phonon bands in a chiral crystal, using the micropolar elasticity theory
for CrNb$_{3}$S$_{6}$. Our main results may be summarized as follows. (I) The
splitting is reached solely within the phonon sector of elementary
excitations, and it is maintained by coupling between the transverse
translational and rotational modes of the micropolar medium. (II) Transverse
acoustic branches of the hybridized phonon spectrum exhibit a ``roton''minimum
reminiscent of elementary excitations in the superfluid helium-4. We argue
that the translational and rotational degrees of freedom of the chiral phonon
system correspond to the Bogoliubov quasiparticles and massive density
fluctuations, respectively, in the superfluid helium-4. In addition, we
discuss acoustic activity originated from nonreciprocity of the phonon
spectrum. {It is expected that the polarization-dependent phonon dipersions demonstrated may lead to chirality-induced cross correlations among lattice, electronic and magnetic degrees of freedom.}

\begin{acknowledgments}
The authors give special thanks to Yusuke Kato for directing our attention to
Ref.\cite{Nozieres,Nozieres2}. We thank Laurence Barron for continuous
encouragement. We also thank Nikolay Baranov, Yoshihiko Togawa and Hiroshi
Yamamoto for stimulating discussions concerning experimental insights. The
authors acknowledge JSPS Bilateral Joint Research Projects (JSPS-RFBR), the
Russian Foundation for Basic Research (RFBR), Grant 20-52-50005. This work was
supported by JSPS KAKENHI Grant Number 17H02923. A.S.O. acknowledges funding
by Act 211 Government of the Russian Federation, Contract No. 02.A03.21.0006,
and the Ministry of Education and Science of Russia, Project No. FEUZ-2020-0054.
\end{acknowledgments}

\end{document}

\title{Supplementary Material for:\\Chirality-induced Phonon Dispersion in a Noncentrosymmetric Micropolar Crystal}
\author{J. Kishine, A. S. Ovchinnikov, and A. A. Tereshchenko}
\maketitle

%\affiliation{$^{1}$ Division of Natural and Environmental Sciences, The Open University
%of Japan, Chiba, 261-8586, Japan}
%\affiliation{$^{2}$ Department of Chemistry, Graduate School of Science, Kyoto
%University, Kyoto 606-8502}
%\affiliation{$^{3}$ Institute for Molecular Science, Okazaki, Aichi 444-8585, Japan}
%\affiliation{$^{4}$ Institute of Natural Sciences and Mathematics, Ural Federal
%University, Ekaterinburg, 620083, Russia}
%\affiliation{$^{5}$ Institute of Metal Physics, Ural Division, Russian Academy of
%Sciences, Ekaterinburg, 620219, Russia}
%\date{\today}

In this Supplementary Material we present technical details related to the
micropolar elasticity theory. In Sec. I, we summarize a general formulation.
In Sec. II, a concrete form of equations of motion is derived for the 622
point group. In Sec. III, the stability condition and a finite gap of the
roton minimum are discussed.

\section{General formulation}

In the linear micropolar continuum, the displacement field vector
$\boldsymbol{u}$ is supplemented by the microrotation field vector
$\boldsymbol{\varphi}$, \textit{which are independent of each other}%
[10,11]. Strain measures of the micropolar media are
described by two tensors, i.e. the microdeformation tensor
\begin{equation}
\varepsilon_{kl}=\partial_{l}u_{k}-\epsilon_{klm}\varphi_{m},
\label{microdeformation}%
\end{equation}
and the wryness (microrotation) tensor
\begin{equation}
\gamma_{kl}=\partial_{l}\varphi_{k},
\end{equation}
where $\epsilon_{klm}$ is the totally antisymmetric tensor (Levi-Civita
symbol) and the Einstein summation convention is used. The microdeformation
tensor $\varepsilon_{kl}$ is even under the iversion operation because
$\partial_{l}$ and $u_{k}$ are polar while $\varphi_{m}$ is axial. On the
other hand, the wryness tensor $\gamma_{kl}$ is odd.

{Note that the microdeformation tensor $\varepsilon_{kl}$ is different
from conventional Cauchy's strain tensor, $\frac{1}{2}\left(  \partial
_{l}u_{k}+\partial_{k}u_{l}\right)  $. To see the difference, it is useful to
note that $\partial_{l}u_{k}$ can be always decomposed into a symmetric and
antisymmetric parts,\begin{equation}
\partial_{l}u_{k}=\underset{\text{Cauchy's strain }\varepsilon_{kl}^{\text{C}}}{\underbrace{\frac{1}{2}\left(  \partial_{l}u_{k}+\partial_{k}u_{l}\right)
}}+\frac{1}{2}\left(  \partial_{l}u_{k}-\partial_{k}u_{l}\right)
=\varepsilon_{kl}^{\text{C}}+\epsilon_{klm}\omega_{m},
\end{equation}
where $\varepsilon_{kl}^{\text{C}}$ denotes the Cauchy's strain tensor and
$\omega_{m}$ does the axial rotation vector. Rewrting this relation as\begin{equation}
\varepsilon_{kl}^{\text{C}}=\partial_{l}u_{k}-\epsilon_{klm}\omega_{m},
\end{equation}
the difference between $\varepsilon_{kl}$ and $\varepsilon_{kl}^{\text{C}}$
becomes clear. In the conventional elasticity theory, the axial rotation
vector, $\boldsymbol{\omega}$, originates from the translational
displacement, $\boldsymbol{u}$, similar to $\varepsilon_{kl}^{\text{C}}$, and describes
rotation around the origin, the Cr-ion position[see Fig.~1(b)].
On the other hand, in the Cosserat (micropolar) theory, the microrotation vector, $\boldsymbol{\varphi}$, is totally independent of $\boldsymbol{u}$.}

The strain energy density is given by the quadratic
form[10,11], $U=\frac{1}{2}A_{klmn}\varepsilon
_{kl}\varepsilon_{mn}+\frac{1}{2}B_{klmn}\gamma_{kl}\gamma_{mn}+C_{klmn}%
\varepsilon_{kl}\gamma_{mn}$. The four-rank tensors $\mathbf{A}$ \ and
$\mathbf{B}$\ are both polar tensors and exist irrespective of inversion
symmetry. They possess the property $A_{klmn}=A_{mnkl}$ and $B_{klmn}%
=B_{mnkl}$, which has an internal symmetry $[\left(  V\right)  ^{2}]^{2}$ in
Jahn's symbol[19]. The pseudotensor $\mathbf{C}$ has an internal
symmetry $V^{4}$ and exists only in crystals without inversion symmetry.
{It provides interaction between the parity-even $\varepsilon_{kl}$
and parity-odd $\gamma_{kl}$. This `translation-rotation' coupling is the
essential feature of chirality as has been highlighted in the Introduction. }
A number of independent elements of the $A_{klmn}$($B_{klmn}$) and $C_{klmn}$
conditioned by permutations of the indices amounts to 45 and 81, respectively.
A further account of the point group symmetry of a given crystal reduces their
number. In addition, the tensor components must maintain positive definiteness
of the internal energy, $U>0$, for all strains to guarantee stability of the
ground state against elementary excitations (phonons). Using the lattice
theory [12], whereby atomic interactions are modelled by springs,
typical magnitudes of the tensor components amount at most to $A_{klmn}%
\sim10^{10}\,\mathrm{{N}\cdot{m}^{-2}}$, $B_{klmn}\sim10^{-10}\,\mathrm{N}$
and $C_{klmn}\sim10^{0}\,\mathrm{{N}\cdot{m}^{-1}}$.

Variation of $U$ with respect to $\mathbf{\varepsilon}_{kl}$ leads to the
linear constitutive equation for the stress tensor, $t_{kl}=\delta
U/\delta\varepsilon_{lk}=A_{klmn}\varepsilon_{mn}+C_{klmn}\gamma_{mn}$.
Similarly, one obtain the couple stress tensor $m_{kl}=\delta U/\delta
\gamma_{kl}=C_{mnlk}\varepsilon_{mn}+B_{lkmn}\gamma_{mn}$.

Now consider an elastic body occupying a region of volume $\mathcal{V}$
bounded by the body surface $\partial\mathcal{V}$. The torque exerted on an
element of the surface area reads a
\begin{equation}
d\boldsymbol{S}=\boldsymbol{\hat{n}}dS
\end{equation}
with the exterior normal $\boldsymbol{\hat{n}}$ to $\partial\mathcal{V}$.

The total torque acting on this surface microelement includes the orbital
torque, $\boldsymbol{x}\times\boldsymbol{T}$, and the spin torque,
$\boldsymbol{M}$, where $\boldsymbol{T}$ denotes a force acting on $dS$. Then,
we have
\begin{equation}
\left(  \boldsymbol{x}\times\boldsymbol{T}+\boldsymbol{M}\right)  _{\alpha
}dS=\epsilon_{\alpha\beta\gamma}x_{\beta}T_{\gamma}dS+M_{\alpha}dS.
\end{equation}
The vectors $\boldsymbol{T}$ and $\boldsymbol{M}$ may be rewritten via the
force stress and coupled stress tensors as
\begin{equation}
T_{\gamma}=t_{\delta\gamma}\hat{n}_{\delta},\text{ \ }M_{\alpha}%
=m_{\delta\alpha}\hat{n}_{\delta}.
\end{equation}

We shall restrict our discussion to a case when there is neither external
volume force nor volume torque. Then, the rate of change of the angular
momentum results from the orbital and spin torques acting on the microelement
surface. This consideration leads to the balance equation,
\begin{align}
\label{BeforeGauss} &  \frac{d}{dt}\underset{\text{volume angular momentum}%
}{\underbrace{\int_{\mathcal{V}}\rho\left(  \overset{\text{orbital}%
}{\overbrace{\epsilon_{\alpha\beta\gamma}x_{\beta}v_{\gamma}}}+\overset
{\text{spin}}{\overbrace{j_{\alpha\beta}\dot{\varphi}_{\beta}}}\right)  dV}%
}\nonumber\\
&  =\underset{\text{surface torque}}{\underbrace{\int_{\partial\mathcal{V}%
}\left(  \overset{\text{orbital}}{\overbrace{\epsilon_{\alpha\beta\gamma
}x_{\beta}t_{\delta\gamma}}}+\overset{\text{spin}}{\overbrace{m_{\delta\alpha
}}}\right)  n_{\delta}dS}},
\end{align}
where $\rho$\ is the mass density and $j_{\alpha\beta}$ is the microinertia tensor.

Applying the Gauss theorem to the r.h.s. of Eq.(\ref{BeforeGauss}) one
obtains
\begin{align}
&  \int_{\mathcal{V}}\rho\left(  \epsilon_{\alpha\beta\gamma}x_{\beta}\dot
{v}_{\gamma}+j_{\alpha\beta}\ddot{\varphi}_{\beta}\right)  dV\nonumber\\
&  =\int_{\mathcal{V}}\partial_{\delta}\left(  \epsilon_{\alpha\beta\gamma
}x_{\beta}t_{\delta\gamma}+m_{\delta\alpha}\right)  dV,
\end{align}
what may be converted to
\begin{align}
&  \int_{\mathcal{V}}\epsilon_{\alpha\beta\gamma}x_{\beta}\left(  \rho\dot
{v}_{\gamma}-\partial_{\delta}t_{\delta\gamma}\right)
dV\nonumber\label{balanceEq}\\
&  =-\int_{\mathcal{V}}\left(  \rho j_{\alpha\beta}\ddot{\varphi}_{\beta
}-\epsilon_{\alpha\beta\gamma}t_{\beta\gamma}-\partial_{\delta}m_{\delta
\alpha}\right)  dV.
\end{align}
Since $\mathcal{V}$\ and $x_{\beta}$ are chosen arbitrarily, the integrands of
both sides in Eq. (\ref{balanceEq}) must be zero. This immediately yields the
equations of motion (EOM),
\begin{equation}
\rho\ddot{u}_{l}=\partial_{k}t_{kl}, \label{BLMom}%
\end{equation}%
\begin{equation}
\rho j_{lk}\ddot{\varphi}_{k}=\partial_{k}m_{kl}+\epsilon_{lmn}t_{mn},
\label{BLAM}%
\end{equation}
for the microdisplacements and microrotations, respectively. Here, $\rho$ is
the mass density and $j_{lk}$ is the microinertia tensor. Being written for
the stran fields only, the EOM read as,
\begin{align}
\rho\ddot{u}_{l}  &  =A_{klmn}\partial_{k}\partial_{m}u_{n}\nonumber\\
&  -\epsilon_{mnr}A_{klmn}\partial_{k}\varphi_{r}+C_{klrn}\partial_{k}%
\partial_{n}\varphi_{r},\label{EOMu}\\
\rho j_{kl}\ddot{\varphi}_{k}  &  =C_{mnlk}\partial_{k}\partial_{m}%
u_{n}-\epsilon_{mnr}C_{mnlk}\partial_{k}\varphi_{r}\nonumber\\
&  +B_{lkmn}\partial_{k}\partial_{n}\varphi_{m}+\epsilon_{lmn}A_{mnpq}%
\partial_{p}u_{q}\nonumber\\
&  -\epsilon_{lmn}\epsilon_{pqr}A_{mnpq}\varphi_{r}+\epsilon_{lmn}%
C_{mnpq}\partial_{q}\varphi_{p}. \label{EOMphi}%
\end{align}
Assuming propagating plane wave solutions, Eqs. (\ref{EOMu},\ref{EOMphi})
provide desired dispersion relations. Furthermore, we focus on the phonon
modes propagating along the chiral axis, namely, the `3'-axis. It means that
spatial derivatives in the above equations are taken with respect to this
coordinate only. Consequently, Eqs. (\ref{EOMu},\ref{EOMphi}) can be recast in
the form
\begin{align}
\rho\ddot{u}_{1}  &  =A_{313n}\partial_{3}^{2}u_{n}-\epsilon_{mnr}%
A_{31mn}\partial_{3}\varphi_{r}+C_{31r3}\partial_{3}^{2}\varphi_{r}%
,\label{EOMu1}\\
\rho\ddot{u}_{2}  &  =A_{323n}\partial_{3}^{2}u_{n}-\epsilon_{mnr}%
A_{32mn}\partial_{3}\varphi_{r}+C_{32r3}\partial_{3}^{2}\varphi_{r},\\
\rho\ddot{u}_{3}  &  =A_{333n}\partial_{3}^{2}u_{n}-\epsilon_{mnr}%
A_{33mn}\partial_{3}\varphi_{r}+C_{33r3}\partial_{3}^{2}\varphi_{r},\\
\rho j_{1k}\ddot{\varphi}_{k}  &  =C_{3n13}\partial_{3}^{2}u_{n}%
-\epsilon_{mnr}C_{mn13}\partial_{3}\varphi_{r}+B_{13m3}\partial_{3}^{2}%
\varphi_{m}\nonumber\\
&  +\epsilon_{1mn}A_{mn3q}\partial_{3}u_{q}-\epsilon_{1mn}\epsilon
_{pqr}A_{mnpq}\varphi_{r}\nonumber\\
&  +\epsilon_{1mn}C_{mnp3}\partial_{3}\varphi_{p},\\
\rho j_{2k}\ddot{\varphi}_{k}  &  =C_{3n23}\partial_{3}^{2}u_{n}%
-\epsilon_{mnr}C_{mn23}\partial_{3}\varphi_{r}+B_{23m3}\partial_{3}^{2}%
\varphi_{m}\nonumber\\
&  +\epsilon_{2mn}A_{mn3q}\partial_{3}u_{q}-\epsilon_{2mn}\epsilon
_{pqr}A_{mnpq}\varphi_{r}\nonumber\\
&  +\epsilon_{2mn}C_{mnp3}\partial_{3}\varphi_{p}\\
\rho j_{3k}\ddot{\varphi}_{k}  &  =C_{3n33}\partial_{3}^{2}u_{n}%
-\epsilon_{mnr}C_{mn33}\partial_{3}\varphi_{r}+B_{33m3}\partial_{3}^{2}%
\varphi_{m}\nonumber\\
&  +\epsilon_{3mn}A_{mn3q}\partial_{3}u_{q}-\epsilon_{3mn}\epsilon
_{pqr}A_{mnpq}\varphi_{r}\nonumber\\
&  +\epsilon_{3mn}C_{mnp3}\partial_{3}\varphi_{p}. \label{EOfi3}%
\end{align}

\section{The point group $622$}

\subsection{Tensor components}

The point group symmetry of a given crystal imposes severe restrictions on the
tensors $\mathbf{A}$, $\mathbf{B}$ and $\mathbf{C}$. For example, in the case
of the $622$ point group, Table 20 of Ref. [20] gives
$\mathbf{A}$\ and $\mathbf{B}$ in the form summarized in Table I. Apparently,
there are only 8 independent elements: $A_{12}$, $A_{13}$, $A_{33}$, $A_{44}$,
$A_{55}$, $A_{66}$, $A_{47}$, and $A_{69}$. {We here used the Sirotin
notation[20] and the index contraction scheme (generalized Voigt
scheme),
\begin{align}
11 &  \rightarrow1,\text{ \ }22\rightarrow2,\text{ \ }33\rightarrow3,\\
23 &  \rightarrow4,\text{ \ }31\rightarrow5,\text{ \ }12\rightarrow
6,\nonumber\\
32 &  \rightarrow7,\text{ \ }13\rightarrow8,\text{ \ }23\rightarrow9.\nonumber
\end{align}
}

\begin{table}[ptbh]%
\begin{tabular}
[c]{c|ccccccccc}
& $11$ & $22$ & $33$ & $23$ & $31$ & $12$ & $32$ & $13$ & $21$\\\hline
$11$ & $A_{11}$ & $A_{12}$ & $A_{13}$ & $0$ & $0$ & $0$ & $0$ & $0$ & $0$\\
$22$ & $A_{12}$ & $A_{11}$ & $A_{13}$ & $0$ & $0$ & $0$ & $0$ & $0$ & $0$\\
$33$ & $A_{13}$ & $A_{13}$ & $A_{33}$ & $0$ & $0$ & $0$ & $0$ & $0$ & $0$\\
$23$ & $0$ & $0$ & $0$ & $A_{44}$ & $0$ & $0$ & $A_{47}$ & $0$ & $0$\\
$31$ & $0$ & $0$ & $0$ & $0$ & $A_{55}$ & $0$ & $0$ & $A_{47}$ & $0$\\
$12$ & $0$ & $0$ & $0$ & $0$ & $0$ & $A_{66}$ & $0$ & $0$ & $A_{69}$\\
$32$ & $0$ & $0$ & $0$ & $A_{47}$ & $0$ & $0$ & $A_{55}$ & $0$ & $0$\\
$13$ & $0$ & $0$ & $0$ & $0$ & $A_{47}$ & $0$ & $0$ & $A_{44}$ & $0$\\
$21$ & $0$ & $0$ & $0$ & $0$ & $0$ & $A_{69}$ & $0$ & $0$ & $A_{66}$%
\end{tabular}
\caption{Nonzero tensor components of $\mathbf{A}$\ and $\mathbf{B}$ fot the
point group 622. There is the constraint $A_{11}=A_{12}+A_{66}+A_{69}$.}%
\end{table}

Table 22 of Ref. [20] gives $\mathbf{C}$ in the form summarized
in Table II. It is readily seen that a number of independent elements equals
to 10: $C_{12}$, $C_{13}$, $C_{31}$, $C_{33}$, $C_{44}$, $C_{47}$, $C_{55}$,
$C_{74}$, $C_{66}$, $C_{69}$.

\begin{table}[ptbh]%
\begin{tabular}
[c]{c|ccccccccc}
& $11$ & $22$ & $33$ & $23$ & $31$ & $12$ & $32$ & $13$ & $21$\\\hline
$11$ & $C_{11}$ & $C_{12}$ & $C_{13}$ & $0$ & $0$ & $0$ & $0$ & $0$ & $0$\\
$22$ & $C_{12}$ & $C_{11}$ & $C_{13}$ & $0$ & $0$ & $0$ & $0$ & $0$ & $0$\\
$33$ & $C_{31}$ & $C_{31}$ & $C_{33}$ & $0$ & $0$ & $0$ & $0$ & $0$ & $0$\\
$23$ & $0$ & $0$ & $0$ & $C_{44}$ & $0$ & $0$ & $C_{47}$ & $0$ & $0$\\
$31$ & $0$ & $0$ & $0$ & $0$ & $C_{55}$ & $0$ & $0$ & $C_{74}$ & $0$\\
$12$ & $0$ & $0$ & $0$ & $0$ & $0$ & $C_{66}$ & $0$ & $0$ & $C_{69}$\\
$32$ & $0$ & $0$ & $0$ & $C_{74}$ & $0$ & $0$ & $C_{55}$ & $0$ & $0$\\
$13$ & $0$ & $0$ & $0$ & $0$ & $C_{47}$ & $0$ & $0$ & $C_{44}$ & $0$\\
$21$ & $0$ & $0$ & $0$ & $0$ & $0$ & $C_{69}$ & $0$ & $0$ & $C_{66}$%
\end{tabular}
\caption{Nonzero tensor components of $\mathbf{C}$\ for the point group 622.
There is the constraint $C_{11}=C_{12}+C_{66}+C_{69}$.}%
\end{table}

\section{Phonon dispersions}

\subsection{Transverse branches}

Making use data from Tables I and II an explicit form of the equations of
motion comes from Eqs. (\ref{EOMu},\ref{EOMphi}). The microinertia tensor can
be written in the diagonal form $\hat{\jmath}=$diag$\left(  j_{1},j_{1}%
,j_{3}\right)  $ by taking into account the almost $3\bar{m}$ ($D_{3d}$) point
group symmetry of the CrS$_{6}$ block. This leads to
\begin{align}
\rho\ddot{u}_{1}  &  =A_{55}\partial_{3}^{2}u_{1}-\left(  A_{55}%
-A_{47}\right)  \partial_{3}\varphi_{2}+C_{74}\partial_{3}^{2}\varphi_{1},\\
\rho\ddot{u}_{2}  &  =A_{55}\partial_{3}^{2}u_{2}-\left(  A_{47}%
-A_{55}\right)  \partial_{3}\varphi_{1}+C_{74}\partial_{3}^{2}\varphi_{2},\\
\rho\ddot{u}_{3}  &  =A_{33}\partial_{3}^{2}u_{3}+C_{33}\partial_{3}%
^{2}\varphi_{3},\\
\rho j_{1}\ddot{\varphi}_{1}  &  =C_{74}\partial_{3}^{2}u_{1}+2\left(
C_{44}-C_{74}\right)  \partial_{3}\varphi_{2}+B_{44}\partial_{3}^{2}%
\varphi_{1}\nonumber\\
&  +\left(  A_{47}-A_{55}\right)  \partial_{3}u_{2}-\left(  A_{44}%
-2A_{47}+A_{55}\right)  \varphi_{1},\\
\rho j_{1}\ddot{\varphi}_{2}  &  =C_{74}\partial_{3}^{2}u_{2}-2\left(
C_{44}-C_{74}\right)  \partial_{3}\varphi_{1}+B_{44}\partial_{3}^{2}%
\varphi_{2}\nonumber\\
&  -\left(  A_{47}-A_{55}\right)  \partial_{3}u_{1}-\left(  A_{44}%
-2A_{47}+A_{55}\right)  \varphi_{2},\\
\rho j_{3}\ddot{\varphi}_{3}  &  =C_{33}\partial_{3}^{2}u_{3}+B_{33}%
\partial_{3}^{2}\varphi_{3}-2\left(  A_{66}-A_{69}\right)  \varphi_{3}.
\end{align}

To simplify a solution of the equations of motion, the linearly polarized
fields are relaced by
\begin{align}
u_{\pm}  &  =u_{1}\pm iu_{2},\\
\varphi_{\pm}  &  =\varphi_{1}\pm i\varphi_{2}%
\end{align}
that are circularly polarized and counterrotating in the $(1, 2)$ plane. The
sign $+$($-$) corresponds to the left (right) circular polarization.

In the chiral basis the EOMs for the transverse modes are
\begin{align}
\rho\ddot{u}_{\pm}  &  =A_{55}\partial_{3}^{2}u_{\pm}+C_{74}\partial_{3}%
^{2}\varphi_{\pm}\mp i\left(  A_{47}-A_{55}\right)  \partial_{3}\varphi_{\pm
},\label{EOMutransverse}\\
\rho j_{\pm}\ddot{\varphi}_{\pm}  &  =C_{74}\partial_{3}^{2}u_{\pm}\mp
i\left(  A_{47}-A_{55}\right)  \partial_{3}u_{\pm}+B_{44}\partial_{3}%
^{2}\varphi_{\pm}\nonumber\\
&  \mp2i\left(  C_{44}-C_{74}\right)  \partial_{3}\varphi_{\pm}-\left(
A_{44}-2A_{47}+A_{55}\right)  \varphi_{\pm}, \label{EOMphitransverse}%
\end{align}
where $j_{+}=j_{-}=j_{1}$.

To get a dispersion relation, we use the plane-wave solutions
\begin{equation}
u_{\alpha}\left(  x_{3},t\right)  =U_{\alpha}e^{-i\left(  \omega
t-kx_{3}\right)  },\text{ }\varphi_{\alpha}\left(  x_{3},t\right)
=\Phi_{\alpha}e^{-i\left(  \omega t-kx_{3}\right)  },
\end{equation}
where the index $\alpha$ labels $+,-$ and $3$.

The EOMs for the transverse modes ($\alpha=\pm$) are written in the matrix
form as
\begin{equation}
\left(
\begin{array}
[c]{cc}%
\rho\omega_{\alpha}^{2}-a_{\alpha} & -\Delta_{\alpha}\\
-\Delta_{\alpha} & \rho j_{\alpha}\omega_{\alpha}^{2}-b_{\alpha}%
\end{array}
\right)  \left(
\begin{array}
[c]{c}%
U_{\alpha}\\
\Phi_{\alpha}%
\end{array}
\right)  =0. \label{PlaneWaveEOM}%
\end{equation}
Here, the $k$-dependent coefficients are introduced
\begin{align}
a_{\pm}  &  =A_{55}k^{2},\\
b_{\pm}  &  =B_{44}k^{2}\mp2\left(  C_{44}-C_{74}\right)  k\nonumber\\
&  +\left(  A_{44}+A_{55}-2A_{47}\right)  ,\\
\Delta_{\pm}  &  =C_{74}k^{2}\pm\left(  A_{55} - A_{47}\right)  k.
\end{align}

It is immediately recognized that the crystal has two branches for each
polarization $\alpha$:\newline(I) the acoustic branch
\begin{equation}
\omega_{\alpha}^{(\text{A})}\left(  k\right)  ^{2}=\frac{1}{2\rho j_{\alpha}%
}\left[  b_{\alpha}+j_{\alpha}a_{\alpha}-\sqrt{\left(  b_{\alpha}-j_{\alpha
}a_{\alpha}\right)  ^{2}+4j_{\alpha}\Delta_{\alpha}^{2}}\right]  ;
\label{AcoustBr}%
\end{equation}
(II) the optical branch
\begin{equation}
\omega_{\alpha}^{(\text{O})}\left(  k\right)  ^{2}=\frac{1}{2\rho j_{\alpha}%
}\left[  b_{\alpha}+j_{\alpha}a_{\alpha}+\sqrt{\left(  b_{\alpha}-j_{\alpha
}a_{\alpha}\right)  ^{2}+4j_{\alpha}\Delta_{\alpha}^{2}}\right]  .
\label{OpticalBr}%
\end{equation}
The acoustic branch starts off at zero frequency, while the optical branch
starts off at some finite value
\begin{equation}
\omega_{\pm}^{(\text{A})}\left(  0\right)  ^{2}=0,\text{ }\omega_{\pm
}^{(\text{O})}\left(  0\right)  ^{2}=\frac{1}{\rho j_{\pm}}\left(
A_{44}+A_{55}-2A_{47}\right)  .
\end{equation}

\subsection{Longitudinal branches}

It is to be noted that the transverse and longitudinal modes are totally
decouple. The EOMs for the longitudinal modes read
\begin{align}
\rho\ddot{u}_{3}  &  =A_{33}\partial_{3}^{2}u_{3}+C_{33}\partial_{3}%
^{2}\varphi_{3},\label{EOMulong}\\
\rho j_{3}\ddot{\varphi}_{3}  &  =C_{33}\partial_{3}^{2}u_{3}+B_{33}%
\partial_{3}^{2}\varphi_{3}-2\left(  A_{66}-A_{69}\right)  \varphi_{3}.
\label{EOMphilong}%
\end{align}
By repeating the procedure presented above one obtains the acoustic and
optical branches given by Eqs. (\ref{AcoustBr},\ref{OpticalBr}), where
$\alpha=3$ and
\begin{align}
a_{3}  &  =A_{33}k^{2},\\
b_{3}  &  =B_{33}k^{2}+2\left(  A_{66}-A_{69}\right)  ,\\
\Delta_{3}  &  =C_{33}k^{2}.
\end{align}
At $k=0$ we have
\begin{equation}
\omega_{3}^{(\text{A})}\left(  0\right)  ^{2}=0,\text{ }\omega_{3}%
^{(\text{O})}\left(  0\right)  ^{2}=\frac{2}{\rho j_{3}}\left(  A_{66}%
-A_{69}\right)  .
\end{equation}

\subsection{Weights of the transverse $u$ and $\varphi$ modes}

The EOMs (\ref{PlaneWaveEOM}) can be recast in the equivalent form
\begin{gather}
\left(
\begin{array}
[c]{cc}%
j_{\alpha}^{1/2}\left(  \rho\omega_{\alpha}^{2}-a_{\alpha}\right)  &
-\Delta_{\alpha}\\
- j_{\alpha}^{1/2} \Delta_{\alpha} & \left(  \rho j_{\alpha}\omega_{\alpha
}^{2}-b_{\alpha}\right)
\end{array}
\right) \nonumber\\
\times\left(
\begin{array}
[c]{c}%
j_{\alpha}^{-1/2}U_{\alpha}\\
\Phi_{\alpha}%
\end{array}
\right)  =0,
\end{gather}
where both $j_{\alpha}^{-1/2}U_{\alpha}$ and $\Phi_{\alpha}$ are
dimensionless, since $j_{\alpha}^{1/2}$ has a length scale of order of the
microelement size.

Inserting in this equation the relation
\begin{equation}
\rho\omega_{\alpha}{}^{2}-a_{\alpha}=\frac{1}{2j_{\alpha}}\left[  b_{\alpha
}-j_{\alpha}a_{\alpha}\mp\sqrt{\left(  b_{\alpha}-j_{\alpha}a_{\alpha}\right)
^{2}+4j_{\alpha}\Delta_{\alpha}^{2}}\right]  ,
\end{equation}
we obtain
\begin{gather}
j_{\alpha}^{-1/2}\left\vert U_{\alpha}\right\vert =\left[  \frac{1}{2}%
+\frac{j_{\alpha}\left(  a_{\alpha}-\rho\omega_{\alpha}^{2}\right)  ^{2}%
}{2\Delta_{\alpha}^{2}}\right]  ^{-1/2}\nonumber\\
=\left[  \frac{1}{2}+\frac{\left(  b_{\alpha} - j_{\alpha}a_{\alpha} \mp
\sqrt{\left(  b_{\alpha}-j_{\alpha}a_{\alpha}\right)  ^{2}+4j_{\alpha}%
\Delta_{\alpha}^{2}} \right)  ^{2}}{8j_{\alpha}\Delta_{\alpha}^{2}}\right]
^{-1/2},
\end{gather}
where the upper/lower sign corresponds to the acoustic/optical branch.

Analogously, we get
\begin{gather}
\left\vert \Phi_{\alpha}\right\vert =\left[  \frac{1}{2}+\frac{\Delta_{\alpha
}^{2}}{2j_{\alpha}\left(  a_{\alpha}-\rho\omega_{\alpha}^{2}\right)  ^{2}%
}\right]  ^{-1/2}\nonumber\\
=\left[  \frac{1}{2}+\frac{2j_{\alpha}\Delta_{\alpha}^{2}}{\left(  b_{\alpha
}-j_{\alpha}a_{\alpha}\mp\sqrt{\left(  b_{\alpha}-j_{\alpha}a_{\alpha}\right)
^{2}+4j_{\alpha}\Delta_{\alpha}^{2}}\right)  ^{2}}\right]  ^{-1/2},
\end{gather}
where the normalization condition
\begin{equation}
\left\vert \frac{j_{\alpha}^{-1/2}U_{\alpha}}{\sqrt{2}}\right\vert
^{2}+\left\vert \frac{\Phi_{\alpha}}{\sqrt{2}}\right\vert ^{2}=1
\end{equation}
is accounted for.

\begin{figure}[ptb]
\begin{center}
\includegraphics[width=75mm]{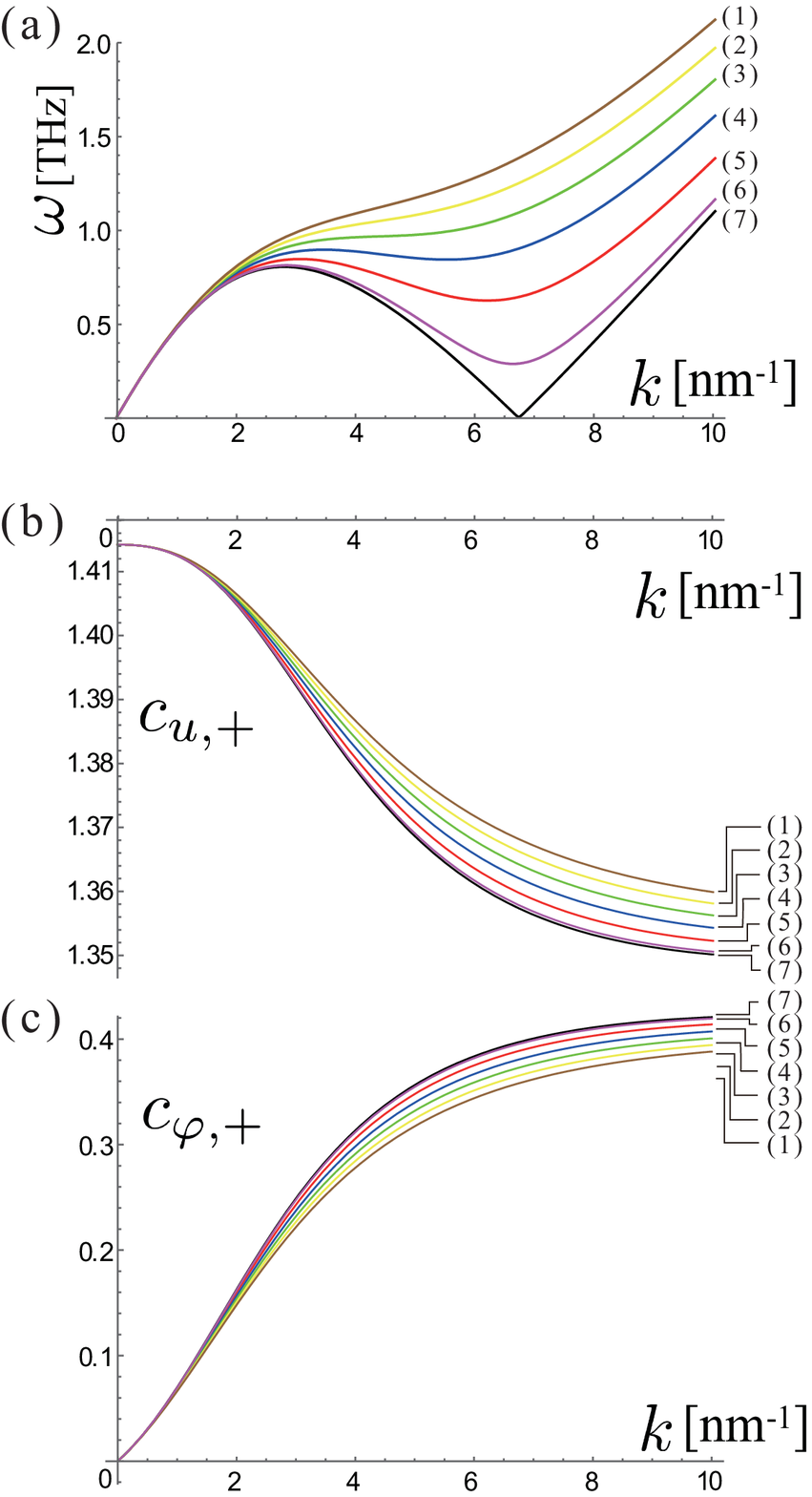}
\end{center}
\caption{Phonon dispersion curves for the chiral micropolar crystal: the
longitudinal acoustic (LA) and optical (LO) branches (green dotted), the
transverse left-handed acoustic (TA-left) and optical (TO-left) branches (blue
solid), the transverse right-handed acoustic (TA-right) and optical
(TO-right)\ branches (red solid). The wavenumber dependence of the weights of
translational and rotational modes in the hybridized TA-right (red solid) and
LA (green dotted) branches. \ $A_{44}=0.2\cdot10^{10}$ N/${\text{m}}^{2}$,
$A_{55}=0.2\cdot10^{10}$ N/${\text{m}}^{2}$, $A_{47}=0.11\cdot10^{10}$
N/${\text{m}}^{2}$, $B_{44}=0.92115\cdot10^{-10}$ N, and $C_{74}=0.35$N/m.
$C_{44}=$(1) $0.35$, (2) $0.36,$ (3) $0.37$, (4) $0.38,$ (5) $0.39,$ (6)
$0.398$, and (7) $0.40$ in the unit of N/m.}%
\label{roton_disp_coeff}%
\end{figure}

In Fig. \ref{roton_disp_coeff}, we show the dispersion curves of the TA-right
branch and corresponding weights. It is clearly seen that the roton minimum is
correlated to the depletion of the $u+$ mode and complementary evolution of
the $\varphi+$ mode.

\section{Stability condition and roton minimum}

Below we formulate the condition of ground state stability against phonon
excitations. For this purpose, we write down the energy $U$ by retaining only
those tensor components which appear in the equations of motion for the case
of the point group 622
\begin{align}
U  &  =\frac{1}{2}\left(  A_{66}+A_{69}\right)  \left(  \varepsilon_{11}%
^{2}+\varepsilon_{22}^{2}\right)  +\frac{1}{2}A_{66}\left(  \varepsilon
_{12}^{2}+\varepsilon_{21}^{2}\right) \nonumber\\
&  +A_{69}\varepsilon_{12}\varepsilon_{21}+\frac{1}{2}A_{33}\varepsilon
_{33}^{2}+C_{33}\varepsilon_{33}\gamma_{33}+\frac{1}{2}B_{33}\gamma_{33}%
^{2}\nonumber\\
&  +\frac{1}{2}A_{44}\left(  \varepsilon_{23}^{2}+\varepsilon_{13}^{2}\right)
+A_{47}\left(  \varepsilon_{32}\varepsilon_{23}+\varepsilon_{31}%
\varepsilon_{13}\right) \nonumber\\
&  +C_{44}\left(  \varepsilon_{23}\gamma_{23}+\varepsilon_{13}\gamma
_{13}\right) \nonumber\\
&  +\frac{1}{2}A_{55}\left(  \varepsilon_{32}^{2}+\varepsilon_{31}^{2}\right)
+C_{74}\left(  \varepsilon_{31}\gamma_{13}+\varepsilon_{32}\gamma_{23}\right)
\nonumber\\
&  +\frac{1}{2}B_{44}\left(  \gamma_{23}^{2}+\gamma_{13}^{2}\right)  ,
\label{Ureduced}%
\end{align}
where we used%
\begin{equation}
A_{11}=A_{66}+A_{69},\text{ \ \ }C_{11}=C_{66}+C_{69}.
\end{equation}

Introducing the vectors
\begin{align}
\boldsymbol{\eta}_{1}  &  =\left(  \varepsilon_{11},\varepsilon_{22}\right)
^{\mathrm{T}},\boldsymbol{\eta}_{2}=\left(  \varepsilon_{12},\varepsilon
_{21}\right)  ^{\mathrm{T}},\boldsymbol{\eta}_{3}=\left(  \varepsilon
_{33},\gamma_{33}\right)  ^{\mathrm{T}},\nonumber\\
\boldsymbol{\eta}_{4}  &  =\left(  \varepsilon_{23},\varepsilon_{32}%
,\gamma_{23}\right)  ^{\mathrm{T}},\boldsymbol{\eta}_{5}=\left(
\varepsilon_{13},\varepsilon_{31},\gamma_{13}\right)  ^{\mathrm{T}},
\end{align}
the stability condition reads
\begin{equation}
U=\frac{1}{2}\sum_{i=1}^{5}\boldsymbol{\eta}_{i}^{\mathrm{T}}\mathbf{\hat{M}%
}_{i}\boldsymbol{\eta}_{i}>0,
\end{equation}
where the matrices are
\begin{align}
\mathbf{\hat{M}}_{1}  &  =\left(
\begin{array}
[c]{cc}%
A_{66}+A_{69} & 0\\
0 & A_{66}+A_{69}%
\end{array}
\right)  ,\\
\mathbf{\hat{M}}_{2}  &  =\left(
\begin{array}
[c]{cc}%
A_{66} & A_{69}\\
A_{69} & A_{66}%
\end{array}
\right)  ,\\
\mathbf{\hat{M}}_{3}  &  =\left(
\begin{array}
[c]{cc}%
A_{33} & C_{33}\\
C_{33} & B_{33}%
\end{array}
\right)  ,\\
\mathbf{\hat{M}}_{4}  &  =\mathbf{\hat{M}}_{5}=\mathbf{\hat{M}}=\left(
\begin{array}
[c]{ccc}%
A_{44} & A_{47} & C_{44}\\
A_{47} & A_{55} & C_{74}\\
C_{44} & C_{74} & B_{44}%
\end{array}
\right)  .
\end{align}
This condition ensures that all eigenvalues of the $\mathbf{\hat{M}}$-matrices
are positive. The terms that contain $\mathbf{\hat{M}}_{4}=\mathbf{\hat{M}%
}_{5}=\mathbf{\hat{M}}$ are of particular importance.

Now, we relate the stability consition with the impossibility of zeroing of
the roton minimum. First, note that the requirement
\begin{equation}
\frac{1}{2}\sum_{\alpha=1,2}\boldsymbol{\eta}_{\alpha}^{T}\mathbf{\hat{M}%
}\boldsymbol{\eta}_{\alpha}>0 \label{U3}%
\end{equation}
with $\boldsymbol{\eta}_{\alpha}^{T}=\left(  \varepsilon_{\alpha3}%
,\varepsilon_{3\alpha},\gamma_{\alpha3}\right)  $, implies $\det
\mathbf{\hat{M}}>0.$ Next, from the dispersion relation (\ref{AcoustBr}) it
can be readily established that the roton minimum touches zero provided
$a_{\alpha}b_{\alpha}=\Delta_{\alpha}^{2}$, where $\alpha=\pm$. This condition
maintains
\begin{gather}
\left(  A_{55}B_{44}-C_{74}^{2}\right)  k^{2}\mp2\left(  A_{55}C_{44}%
-A_{47}C_{74}\right)  k\nonumber\\
+A_{44}A_{55}-A_{47}^{2}=0. \label{zeroingofroton}%
\end{gather}
It is straightforward to show that
\begin{equation}
D/4=-A_{55}\det\mathbf{\hat{M}}. \label{D4}%
\end{equation}
Because $A_{55}$ must be positive, the stability condition, $\det
\mathbf{\hat{M}}>0$, leads to $D<0$, i.e., Eq. (\ref{zeroingofroton}) has no
nontrivial solutions for $k$. Therefore, the roton minimum never touches zero
and a "roton" instability of the crystalline order never occurs.